\documentclass[aps,prl,twocolumn,a4paper,showpacs,superscriptaddress,floatfix,10pt]{revtex4-1}

\usepackage{amsthm,amsmath,amsfonts,amssymb,verbatim, color}
\usepackage{graphicx}
\usepackage{subfigure}
\usepackage{epsfig,slashed}
\usepackage{helvet}
\usepackage{epstopdf}
\usepackage{bbold}
\usepackage{cancel}
\usepackage{color}
\usepackage[colorlinks=true,citecolor=blue,
linkcolor=blue,urlcolor=blue]{hyperref}
\newcommand{\ket}[1]{\ensuremath{\left| #1 \right\rangle}}
\newcommand{\bra}[1]{\ensuremath{\left\langle #1 \right|}}

\def\righta{\rightarrow}

\begin{document}

\title{Classical-quantum mixing in the random 2-satisfiability problem}

\author{Ionut-Dragos Potirniche}
\affiliation{Department of Physics, University of California, Berkeley, California, 94720-7300, USA}
\affiliation{Department of Physics, Princeton University, Princeton, New Jersey 08544, USA}

\author{C. R. Laumann}
\affiliation{Department of Physics, University of Washington, Seattle, Washington 98195, USA}
\affiliation{Max-Planck-Institut f\"{u}r Physik Komplexer Systeme, 01187 Dresden, Germany}

\author{S. L. Sondhi}
\affiliation{Department of Physics, Princeton University, Princeton, New Jersey 08544, USA}
\affiliation{Max-Planck-Institut f\"{u}r Physik Komplexer Systeme, 01187 Dresden, Germany}

\date\today

\begin{abstract}
Classical satisfiability (SAT) and quantum satisfiability (QSAT) are complete problems for the complexity classes NP and QMA, respectively, and are believed to be intractable for both classical and quantum computers.
Statistical ensembles of instances of these problems have been studied previously in an attempt to elucidate their typical, as opposed to worst-case, behavior. 
In this paper, we introduce a new statistical ensemble that interpolates between classical and quantum. 
For the simplest 2-SAT--2-QSAT ensemble, we find the exact boundary that separates SAT and UNSAT instances.
We do so by establishing coincident lower and upper bounds, in the limit of large instances, on the extent of the UNSAT and SAT regions, respectively.
\end{abstract}


\maketitle

The potential power of quantum computers drives the immense effort to build and understand them.
There are two primary theoretical approaches to characterizing the precise extent of this potential.
Complexity theory \citep{Arora} proceeds by identifying so-called `complete' problems which are the hardest problems in a given class, such as NP.
Classically, under the widely believed conjecture that P $\neq$ NP, \emph{all} algorithmic approaches to NP-complete problems such as satisfiability (SAT) must fail on at least some subset of worst-case instances.
Over the last decade, these venerable considerations have been extended to the quantum case, where QMA-complete problems are now believed to be intractable and to capture the intrinsic differences between quantum and classical computing \citep{kitaevbook}.
The natural quantum generalization of SAT, so-called quantum satisfiability (QSAT) \citep{bravyi,gosset}, is conveniently such a QMA$_1$-complete problem \footnote{While
both NP and QMA complete problems are believed to be hard for both classical and
quantum computers to solve, the latter are such that a classical computer cannot
even verify the solution efficiently}.

In the second approach, the introduction of an appropriate measure on the instances of a problem generates a question in statistical Physics.
Instead of worrying about worst-case complexity, we attempt to understand the structure of problems that are typical with respect to the measure.
For example, phase transitions, which arise as functions of parameters controlling the measure, often signal the regimes where the most complicated problems may be found \citep{weigt}.
This approach builds on the seminal insight of Fu and Anderson \citep{anderson} that the intractability of NP-complete problems is a form, indeed an extreme one, of (spin) glassiness. Ensembles of both classical SAT \cite{montanaribook,kirkpatrick2} and, more recently, QSAT \cite{laumann2, laumann3, hsu} have been studied in this fashion.

In this paper we build on the second approach and introduce new ensembles that interpolate between the SAT and QSAT ones. The mixture provides a convenient framework for characterizing the crossover from classical to quantum search complexity. For example, the classical PCP theorem \citep{PCPthm} shows that it is computationally hard to approximately determine the ground state of the SAT problem, while it is still an open question whether an analogous hardness result applies in the quantum case. 
The interpolation allows the study of the crossover in entanglement properties of low-energy states, which may bear on this question. Similarly, in statistical physics, the mixed ensemble can shed light on entanglement phase transitions in spin-glass models with quenched disorder: the classical problem has no entangled solutions, whereas the mixed one exhibits entangled states in the UNSAT regime.

\begin{figure}[h]
\centering
\includegraphics[width=\columnwidth]{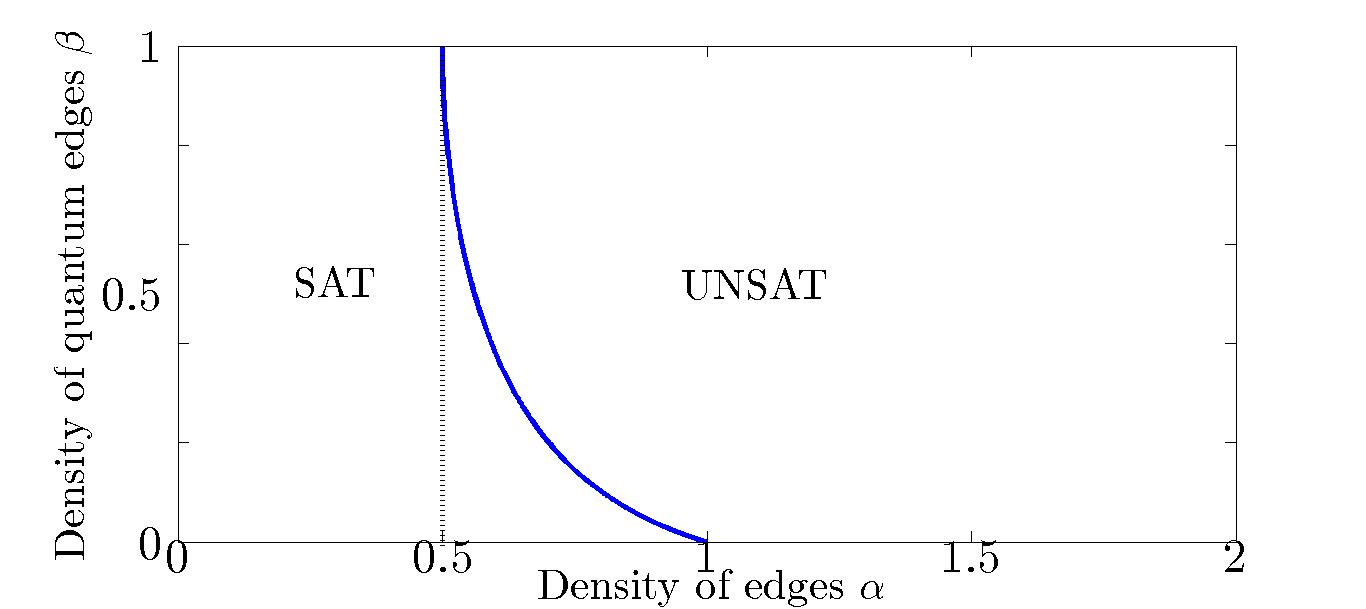}
\caption{The phase boundary in the $\alpha-\beta$ plane that separates the SAT and the UNSAT regimes for the mixed classical-quantum problem. The dashed line indicates the emergence of a giant component in the ER graph (the percolation phase transition). $\beta=0$ corresponds to the 2-SAT problem and $\beta=1$ corresponds to the 2-QSAT problem.}
\label{fig:phasediag}
\end{figure}

Specifically for the 2-SAT--2-QSAT interpolation, we show that there is a sharp phase boundary (Fig.~\ref{fig:phasediag}) that we determine rigorously by deriving coincident lower and upper bounds on the extent of the UNSAT and SAT regions, respectively. The interest in this interpolation flows considerably from a ``geometrization'' theorem that applies to the QSAT limit \citep{laumann2}. As this result is not widely known, we begin with a quick review of the relevant background, which will also enable a proper definition of the problem studied herein. We then present our central technical results on the phase boundary, and we close with some remarks on the lessons learned and future directions.

\noindent
\textbf{SAT, QSAT, and the mixed ensemble:} An instance of $k$-QSAT is defined by a positive-semidefinite Hamiltonian on $N$ qubits given by the sum of $M$ $k$-body interactions, $H = \sum_{m=1}^M \Pi_m$. 
Here, each interaction $\Pi_m = \ket{\phi}\bra{\phi}_{m}$ projects onto a particular state in the local Hilbert space of the $k$ qubits associated with interaction $m$.
The computational problem is to decide whether $H$ has a ground state of energy zero. 
If so, the problem is SAT, and if not it is UNSAT \footnote{The ground state energy is at least a promised gap $\Delta$ above zero if it is not precisely zero.}.
If the states $\ket{\phi}_{m}$ are computational basis states, we recover classical $k$-SAT as a special case. 
We can now summarize the statistical ensembles used in previous work. 
These involve the uniform Erd\H{o}s-R\'{e}nyi measure  \cite{hypergraphs,erdos,bollobas}, parametrized by the `clause density' $\alpha = M/N$, over the set of $k$-hypergraphs representing the interactions in $H$. 
The interaction associated with each hyperedge is likewise uniformly chosen  from the $2^k$ projectors in classical SAT or from the Haar measure over the rays in the $2^k$-dimensional Hilbert space for QSAT. 
We observe that the latter makes classical SAT instances highly non-generic within the QSAT ensemble, and so these two can be expected to behave very differently, and indeed they do \citep{montanaribook, kirkpatrick2}.

The classical ensemble has been studied intensively. 
The broad picture that has emerged is that for all $k \ge 2$ there is a sharp SAT-UNSAT transition in the $N \rightarrow \infty$ limit as a function
of $\alpha$. 
For $k \ge 3$, there are additional transitions in the SAT regime wherein the structure of the solution space changes. 
For the quantum ensemble, it is known that there is a SAT-UNSAT transition for $k \ge 2$ \cite{laumann2,laumann3,bravyimoore,beaudrap} and that there is at least one sharp transition involving the growth of entanglement in the satisfying states for $k \ge 12$ \cite{laumann3,Ambainis:2012aa}. 
A remarkable result to come out of the quantum generalization is a ``geometrization theorem'' \citep{laumann2} wherein uniformly chosen
quantum projectors on {\it any} graph exhibit the same dimension of the satisfying manifold with probability 1. 
This reduces the generic QSAT decision problem to a purely graph-theoretic question! 
The identification of this implicit graph-theoretic property for $k \ge 3$ and understanding its computational difficulty is an outstanding problem. 

As advertised above, we initiate a new approach by introducing ensembles that interpolate between the fully classical and quantum regimes. 
We do this by constraining each realization to include a fraction $\beta$ of uniformly chosen quantum projectors and $1 -\beta$ uniformly chosen classical ones. 
As $\beta$ varies between 0 and 1, we pass from the classical ensemble to the quantum one. 
As the classical ensemble does not exhibit geometrization and typically has larger satisfying manifolds, the interpolation has the potential to
shed light on the emergence of geometrization in the quantum limit and thence on precisely where in the ensemble one might look for genuinely difficult quantum cases.

As a first step in this program, we study the case of 2-SAT/QSAT. We generate the underlying 2-graphs
by drawing edges between any two sites with probability $\alpha N / \binom{N}{2}$. In the thermodynamic limit $N \righta \infty$, this generates an Erd\H{o}s-R\'{e}nyi (ER) random graph with $M = \alpha N$ expected edges. For each edge $m = 1...M$ we label it `quantum' with probability $\beta$ and `classical' with probability $1-\beta$: we write $e_{m} \in \{Q,C\}$ if the edge 
between sites $(m,m+1)$ is  quantum or classical. The purely classical and quantum limits are very well understood. At $\beta =0$ \cite{goerdt, friedgut, bollobascaling} there is a sharp SAT-UNSAT transition at $\alpha_{c} = 1$, while for $\beta = 1$ there is a SAT-UNSAT transition at $\alpha_{q} = 1/2$ \citep{laumann2}. The quantum transition coincides with the emergence of a giant component in the underlying random graph \citep{erdos}.

\noindent
\textbf{The snip-core:} The primary tool in our analysis is snipping qubits out of an interaction graph $G$. 
Classically, a node $i$ such that $G = G' \cup \{i\}$ is \emph{snippable} if all of the bonds connected to it agree about the bit arrangement they locally disfavor.
All clauses attached to such a snippable node $i$ can then be trivially satisfied by assigning the appropriate value to qubit $i$ without reference to the state on G'. 
Thus, these bonds can be snipped from the graph $G$ to produce a smaller graph $G'$, which is SAT if and only if the original $G$ is. 

This definition extends naturally to the mixed classical-quantum problem ($\beta \neq 0$): a degree-1 site $i$ with a quantum projector attaching it to site $j$ of $G'$ is snippable. From Bravyi's construction \citep{bravyi}, we know that the satisfying state for $G'$ can be written as a product state $\ket{\psi_{j}} \otimes \ket{\Psi_{G' \backslash \{j\}}}$. If the quantum edge attaching site $i$ disfavors the state $\ket{\phi_{j,i}}$, then we can use the transfer matrix $T_{\phi} = \epsilon \phi^{\dagger}_{j,i}$ ($\epsilon$ is the $2 \times 2$ Levi-Civita symbol) to find $\ket{\xi_{i}} = T_{\phi} \ket{\psi_{j}}$ such that $\bra{\phi_{j,i}} \ket{\psi_{j}} \otimes \ket{\xi_{i}} = 0$. Therefore, $\ket{\xi_{i}} \otimes \ket{\psi_{j}} \otimes \ket{\Psi_{G' \backslash \{j\}}}$ is a satisfying state for $G = G' \cup \{i\}$, which shows that a degree-1 site with a quantum projector attached is snippable.

However, if a generic quantum edge attaches to a site $i$ of degree greater than 1, it cannot be locally satisfied by the state on $i$ without reference to the rest of the graph (with probability 1). Thus, any site of degree at least 2 with a quantum edge attached is unsnippable.

For a random instance $G$ of the mixed problem, we can iteratively remove snippable sites and the incident edges in a similar fashion to the ``leaf removal'' algorithm \citep{mezard}. When there are no snippable qubits left, this algorithm stops and we end up with a unique maximal \emph{snip-core}. Clearly, $G$ is SAT iff its snip-core is SAT. Moreover, if the snip-core is empty, then $G$ is SAT. Behind our considerations lies the result that when $G$ is SAT, we can always find a satisfying product state \citep{bravyi}. We will now analyze the structure and probability of non-empty snip-cores in order to determine the SAT-UNSAT boundary.

\noindent
{\bf Upper bound on SAT region:} The basic idea in this part is to identify an UNSAT motif that must be present on 
all snip-cores for $\alpha > \alpha_c(\beta)$, thus establishing an upper bound on the extent of the SAT region. Toward that end, we first note that
the simplest motif that is unsnippable is the \emph{unsnippable loop}. A loop with classical and quantum edges is said to be \emph{unsnippable} if all of its sites are unsnippable [see Fig.~2(a)]. We can find a simple example in the fully classical problem: a loop that dislikes $01$ on each edge. While this loop is unsnippable, it is SAT: it has exactly two satisfying states comprised of all sites 0 or all sites 1. The same is true of the other $2^{L}$ such classical unsnippable loops of length $L$ that are equivalent under ``gauge'' transformations. It can be shown that the fully quantum loop also has two SAT states \citep{laumann2, bravyi}. Moreover, based on the Geometrization Theorem, we can conclude that the dimension of the kernel of the fully quantum problem is a lower bound for the mixed problem, whereas the fully classical one is an upper bound: we start from the classical unsnippable loop and slowly turn classical projectors into generic quantum ones; this can only decrease the degeneracy of satisfying states. But since both kernels have dimension 2, we conclude that the mixed classical-quantum unsnippable loop always has exactly two linearly independent SAT states. Finally, an unsnippable but UNSAT motif can be constructed by decorating
the unsnippable loop with two unsnippable cross-links---strings joining two different points on the loop whose interior sites
are unsnippable [Fig.~2(b)] that penalize the two SAT states.

We now turn to the probability of finding UNSAT unsnippable loops with cross-links. Quite generally, the expected number of subgraphs 
$A$ in the ER ensemble on $N$ nodes is
\begin{equation}
\mathbb{E}(\# \;\mathrm{of}\; A) = \frac{N!}{(N - |A|)! \mathrm{Aut}(A)}p^{e(A)},
\end{equation}
where $|A|$ and $e(A)$ represent the number of vertices and edges of $A$, respectively; $\mathrm{Aut}(A)$ is the number of automorphisms of $A$ and $p = \alpha N / \binom{N}{2}$. For a loop of length $L$ we have $|L| = e(L) = L$ and $\mathrm{Aut}(L) = 2L$. Introducing the probability that a given loop is unsnippable, we find the expected number of unsnippable
loops  $\#_{\mathrm{uns}}(L)$ is
\begin{equation}\label{eq:numberofloops}
\#_{\mathrm{uns}}(L) = \binom{N}{L} \frac{L!}{2L} \left(\frac{2\alpha}{N-1} \right)^{L} p(L \; \mathrm{is} \; \mathrm{unsnippable}).
\end{equation}
For any fixed length $L$ and as $N \rightarrow \infty$, $\#_{\mathrm{uns}}(L)$ scales as $\mathcal{O}(N^{0})$. More generally, the number of subgraphs with $m = e(A) - |A|$ cross-links vanishes in the thermodynamic limit as $\mathcal{O}(N^{-m})$. This result holds irrespective of the form of $ p(A \; \mathrm{is} \; 
\mathrm{unsnippable})$ since this probability has no explicit dependence on $N$. 
It follows that in order to get a non-vanishing number of UNSAT motifs we must consider \emph{giant loops} whose size scales with $N$ as $L = lN$ ($0<l \leq 1$). 

To calculate the number of unsnippable giant loops we need the last factor in Eq.~(\ref{eq:numberofloops}). For a collection $\{e_{i}\}$ of $M$ edges, the probability that site $i$ is unsnippable is $1 - \delta_{e_{i},C}\delta_{e_{i+1},C}/2$. In words, if it is connected to a quantum edge then it is unsnippable with probability 1; otherwise it is unsnippable with probability $1/2$. Also, from the definition of the random ensemble we have $p(e_{i} = Q) = \beta$ and $p(e_{i}=C) = 1-\beta$. Hence, for a loop of length $L$
\begin{equation}\label{eq:pforcing}
p(\mathrm{L \; is\; unsnippable}) = \sum_{\{e_{i}\}} \prod_{i=1}^{L} p(e_{i})\left(1 - \frac{\delta_{e_{i},C}\delta_{e_{i+1},C}}{2}\right).
\end{equation}
Using a standard transfer matrix technique and focusing solely on the dominant eigenvalue $\lambda_{+}$ which controls the
result for large loops, we find that
\begin{equation}
p(\mathrm{L \; is\; unsnippable}) \approx \left( \frac{1+\beta + \sqrt{-7\beta^{2}+10\beta+1}}{4} \right)^{L}.
\end{equation}

With this in hand, we return to Eq.~(\ref{eq:numberofloops}) and use Stirling's approximation to find the extensive part of the entropy,
\begin{equation}\label{eq:extensiveentropy}
S_{\mathrm{uns}}(l) = N \left[l\left(\log(2\alpha \lambda_{+})-1   \right) - (1-l)\log (1-l) \right].
\end{equation}
Giant unsnippable loops proliferate exponentially in $N$ if the entropy function is positive for some $1\geq l>0$. Since $S_{\mathrm{uns}}(0) = 0$ and $S^{'}_{\mathrm{uns}}(l) = N \left[ \log(2\alpha \lambda_{+}) + \log(1-l) \right] \leq N\log(2\alpha \lambda_{+})$, we see that $S_{\mathrm{uns}}(l)$ is a negative and decreasing function on $0 < l \leq 1$ for $2\alpha \lambda_{+} < 1$. However, for $2\alpha\lambda_{+} >1$, the entropy goes positive for small $l$ and large
numbers of giant unsnippable loops emerge. As we find $e^{\mathcal{O}(N)}$ loops on just $N$ sites, the loops must
intersect and overlap repeatedly. It follows that any given giant loop is covered by a finite density of cross-links and therefore the probability of finding an UNSAT unsnippable loop with cross-links approaches
1 as $N \rightarrow \infty$ for $\alpha$ greater than
\begin{equation}\label{eq:upperbound}
\alpha_{c}(\beta) = \frac{2}{1+\beta + \sqrt{-7\beta^2 + 10\beta + 1}}.
\end{equation}
Hence, we conclude that that $\alpha_c(\beta)$ is an upper bound on the extent of the SAT region at
any fixed $\beta$. 

\noindent
{\bf Lower bound on UNSAT region:} We will now show that for $\alpha < \alpha_{c}(\beta)$ the snip-core is always SAT
so that  $\alpha_{c}(\beta)$ is also a lower bound on the extent of the UNSAT region. We do this by showing that the
snip-core must contain one of a finite list of motifs \footnote{We use motif to mean a fixed subgraph up to isomorphisms, irrespective of its frequency of occurrence. This is midway between the everyday meaning of the term and the more technical graph theoretic usage which makes reference to its frequency} and that all of these {\it except} SAT unsnippable loops are not present as
$N \rightarrow \infty$ for $\alpha < \alpha_{c}(\beta)$.

To show that there is a finite list of necessary motifs, we start from an arbitrary site in the snip-core which we label $1$  and walk along any edge which we label $e_{1}$ that connects it to a site $2$ [see Fig.~2(c)]. At step 2, we walk along
an edge $e_2$ to site $3$ such that $2$ is unsnippable with respect to edges $e_1$ and $e_2$. Thus, $e_{2}$ can either be quantum or, if it is classical, it has to disagree with $e_{1}$ on site 2.
From this point onward, we take further steps as follows:
\begin{itemize}
\item At step $k$: we move along an edge $e_{k}$ that connects it to $k+1$ such that site $k$ is unsnippable. 
\item Iterate until we pass through a site $i$ twice. 
\end{itemize}
Since the size of the snip-core is finite at any given N, then each such path must be self-intersecting. Therefore, the algorithm stops in either of the following scenarios:
\begin{enumerate}
\item The path returns to the starting point ($i=1$) and we end up with a loop. There are two subcases:
	\begin{enumerate}
		\item If all sites on the loop have degree 2 then the loop must be unsnippable [Fig.~2(a)].
		\item If there exists a site on the loop that has a degree of at least 3 then we walk one step away from the loop 
	starting at that site as before and get a ``lasso motif'' [Fig.~2(c)].
	\end{enumerate}
\item The path crosses itself at a site $i \neq 1$ and we encounter the same lasso motif.
\end{enumerate}
At this point, we continue from the open end of the lasso and generate unsnippable sites as before. Again, our path necessarily returns and touches the lasso. When this happens, we end up with one of three motifs: an unsnippable loop with one cross-link [Fig.~2(d)], a ``figure eight'' [Fig.~2(e)] or  a ``dumbbell'' [Fig.~2(f)]. Together with the unsnippable loop [Fig.~2(a)], these constitute the set of structures of which at least one {\it must} be present on each non-empty disconnected component of a snip-core. 
Strictly speaking, we should classify somewhat more finely by specifying the unsnippability of each loop passing through the
degree 3 and 4 sites in these motifs. But that only changes our estimates below by constant factors as $N \rightarrow \infty$,
so we refrain from exhibiting these details here.

Now for the frequency of occurrence of these motifs. In the limit $N \rightarrow \infty$ for any fixed number of sites $L$, the expected number of loops is $O(N^0)$, while the expected numbers of the other three are $O(1/N)$ and hence vanish. As before,
we are led to examine giant versions of these graphs. The most optimistic case assumes that all the individual legs of the motifs are large and diverge with $N$. In this case, the expected numbers take the form
\begin{equation}\label{eq:numberdumbbells}
\mathbb{E} (\# \; \mathrm{motifs}) =   c \binom{N}{L-1} \frac{(L-1)!}{a} \left(\frac{2\alpha}{N-1}\right)^{L} \lambda_{+}^{L}, 
\end{equation}
where $a$ is the number of automorphisms ($a = 4$ for the figure eight and the dumbbell and $a=2$ for the loop with a single cross-link) and $c$ is an $\mathcal{O}(1)$ number (dependent on $\beta$) associated with the precise unsnippability of vertices that have degree at least 3 alluded to above.

For $\alpha < \alpha_c (\beta)$ these expected numbers
vanish also for large motifs of size $L = l N^{\gamma}$ as $N \rightarrow \infty$ for any $l$ and $0<\gamma \leq 1$. Using Stirling's approximation, the $\mathcal{O}(N^{\gamma})$ part of the entropy $S(l) = \log(\mathbb{E} (\# \; \mathrm{motifs}))$ is approximately
\begin{equation}
S(l) = lN^{\gamma} \log(2\alpha\lambda_{+}) - lN^{\gamma} - N(1-lN^{\gamma-1})\log(1-lN^{\gamma-1}).
\end{equation}
Once again, $S(0) = 0$ and the derivative $S'(l) = N^{\gamma} \left[\log(2\alpha \lambda_{+}) + \log(1-lN^{\gamma-1}) \right] \leq N^{\gamma} \log(2\alpha\lambda_{+})$ for any $l$. We see that for $\alpha< \alpha_{c}(\beta)$, $S(l)$ is a negative and decreasing function so the expected number of motifs vanishes in the thermodynamic limit as $e^{-\mathcal{O}(N^{\gamma})}$.

Hence, the loop with cross-link, figure eight, and dumbbell are entirely absent for $\alpha < \alpha_c (\beta)$ and the snip-core is either empty or composed entirely
of unsnippable loops. Such snip-cores are SAT so we can conclude that our starting graphs are SAT for  $\alpha < \alpha_c (\beta)$ with probability 1 and thus $\alpha_c(\beta)$ is a lower bound on the extent of the UNSAT region at fixed $\beta$.  Putting together the upper and the lower bounds, we conclude that $\alpha_{c}(\beta)$ from Eq.~(\ref{eq:upperbound}) represents the exact location of the phase boundary between SAT and UNSAT in the $\alpha-\beta$ plane, as shown in Fig.~\ref{fig:phasediag}. 

\begin{figure}[t]
\centering
\includegraphics[width=\columnwidth]{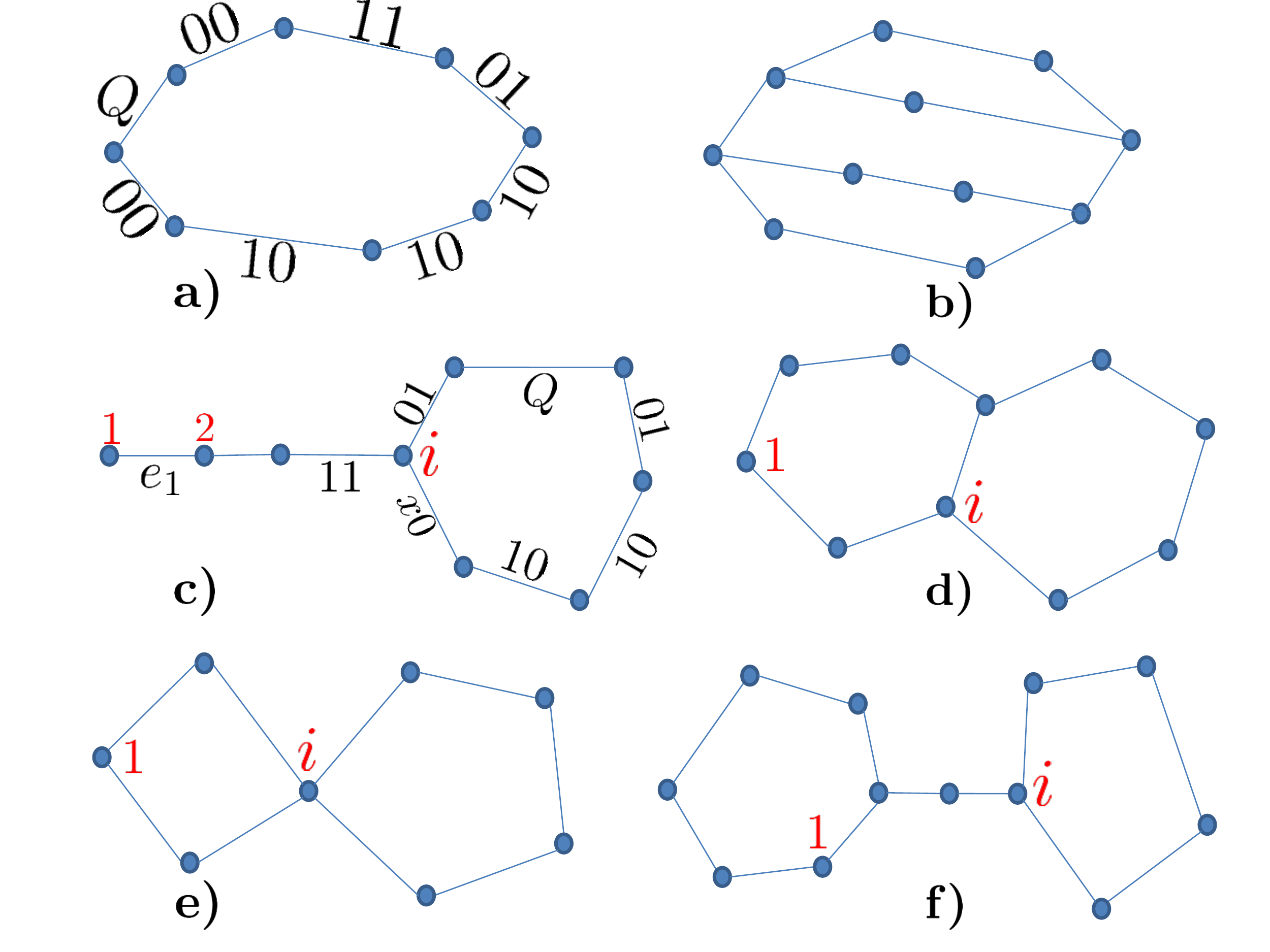}
\caption{\\
(a) An unsnippable loop that contains both classical and quantum edges.\\
(b) A loop with two cross-links. If all the strings are unsnippable and the cross-links penalize the loop's two linearly independent satisfying states, this motif becomes UNSAT.\\
(c) The ``lasso'' motif. We start from a random site $1$ and move along an unsnippable path until we pass through the same site $i$ twice. The loop can either be unsnippable ($x=1$) or snippable at site $i$ ($x=0$).\\
(d) The loop with a single cross-link. We encounter this motif if we traverse the dangling branch of the lasso and end up at a site located on the initial loop.\\
(e) The ``figure eight.'' We encounter this motif if we traverse the dangling branch of the lasso and end up at the same starting point $i$.\\
(f) The ``dumbbell.'' We encounter this motif if we traverse the dangling branch of the lasso and pass through the same point twice (located outside of the initial loop).}
\label{fig:graphs}
\end{figure}

\noindent
{\bf Concluding remarks:} In the preceding, we introduced a new family of mixed classical-quantum ensembles for the
$k$-SAT/QSAT problems, and we established the exact phase diagram for the simplest member of this family with $k=2$. We note that
the shape of the phase boundary is consistent with what is known about the limits. The quantum limit is insensitive to the
choice of projectors, and we find that the restriction of a dilute concentration of projectors to classical values barely shifts 
the phase boundary. The classical limit is sensitive to the choice of projectors, and the phase boundary near it is maximally
sensitive to the inclusion of a dilute set of quantum projectors. A problem that we leave open is the nature of scaling near
the phase boundary. One basic question concerns the scaling of the probability $P_{\mathrm{SAT}}$ that a random instance is 
SAT. Work on the classical problem by Bollob\'{a}s \emph{et al.} \citep{bollobascaling} showed that this has the scaling form
$P_{\mathrm{SAT}} (N, \alpha) = f((\alpha-\alpha_c)/N^{1/3})$. Work on the ER ensemble by Bollob\'{a}s \citep{bollobasevolution} and {\L}uczak \citep{luczak} implies that the same scaling holds for the purely quantum problem. Qualitatively, our work
makes sense of this apparent coincidence and implies that this scaling will hold everywhere along the phase boundary---that everywhere there exists a single transition involving the  proliferation of unsnippable loops with the mix of projectors along such loops changing continuously with $\beta$. However, more work is required to show this rigorously and to try and extract more
detailed information on the scaling functions near the transition. Finally, we intend to examine similar ensembles for $k=3$. 
The current work suggests that perturbing about the classical limit by introducing a dilute set of quantum projectors into
the Hamiltonian could be informative in quantum mechanical perturbation theory.

\noindent
{\bf Acknowledgments:} We are grateful to M. Bardoscia for useful discussions, and to A. Scardicchio and R. Moessner for collaboration on related work. This work was supported by The John Templeton Foundation, the Alexander von Humboldt-Stiftung, and the Deutsche Forschungsgemeinschaft via the Gottfried Wilhelm Leibniz Prize Programme at MPI-PKS (SLS).

\bibliography{biblio}

%merlin.mbs apsrev4-1.bst 2010-07-25 4.21a (PWD, AO, DPC) hacked
%Control: key (0)
%Control: author (72) initials jnrlst
%Control: editor formatted (1) identically to author
%Control: production of article title (-1) disabled
%Control: page (0) single
%Control: year (1) truncated
%Control: production of eprint (0) enabled
\begin{thebibliography}{27}%
\makeatletter
\providecommand \@ifxundefined [1]{%
 \@ifx{#1\undefined}
}%
\providecommand \@ifnum [1]{%
 \ifnum #1\expandafter \@firstoftwo
 \else \expandafter \@secondoftwo
 \fi
}%
\providecommand \@ifx [1]{%
 \ifx #1\expandafter \@firstoftwo
 \else \expandafter \@secondoftwo
 \fi
}%
\providecommand \natexlab [1]{#1}%
\providecommand \enquote  [1]{``#1''}%
\providecommand \bibnamefont  [1]{#1}%
\providecommand \bibfnamefont [1]{#1}%
\providecommand \citenamefont [1]{#1}%
\providecommand \href@noop [0]{\@secondoftwo}%
\providecommand \href [0]{\begingroup \@sanitize@url \@href}%
\providecommand \@href[1]{\@@startlink{#1}\@@href}%
\providecommand \@@href[1]{\endgroup#1\@@endlink}%
\providecommand \@sanitize@url [0]{\catcode `\\12\catcode `\$12\catcode
  `\&12\catcode `\#12\catcode `\^12\catcode `\_12\catcode `\%12\relax}%
\providecommand \@@startlink[1]{}%
\providecommand \@@endlink[0]{}%
\providecommand \url  [0]{\begingroup\@sanitize@url \@url }%
\providecommand \@url [1]{\endgroup\@href {#1}{\urlprefix }}%
\providecommand \urlprefix  [0]{URL }%
\providecommand \Eprint [0]{\href }%
\providecommand \doibase [0]{http://dx.doi.org/}%
\providecommand \selectlanguage [0]{\@gobble}%
\providecommand \bibinfo  [0]{\@secondoftwo}%
\providecommand \bibfield  [0]{\@secondoftwo}%
\providecommand \translation [1]{[#1]}%
\providecommand \BibitemOpen [0]{}%
\providecommand \bibitemStop [0]{}%
\providecommand \bibitemNoStop [0]{.\EOS\space}%
\providecommand \EOS [0]{\spacefactor3000\relax}%
\providecommand \BibitemShut  [1]{\csname bibitem#1\endcsname}%
\let\auto@bib@innerbib\@empty
%</preamble>
\bibitem [{\citenamefont {{Arora}}\ and\ \citenamefont
  {{Barak}}(2009)}]{Arora}%
  \BibitemOpen
  \bibfield  {author} {\bibinfo {author} {\bibfnamefont {S.}~\bibnamefont
  {{Arora}}}\ and\ \bibinfo {author} {\bibfnamefont {B.}~\bibnamefont
  {{Barak}}},\ }\href@noop {} {\emph {\bibinfo {title} {Complexity Theory: A
  Modern Approach}}}\ (\bibinfo  {publisher} {Cambridge University Press},\
  \bibinfo {address} {Cambridge},\ \bibinfo {year} {2009})\BibitemShut
  {NoStop}%
\bibitem [{\citenamefont {{Yu}}\ \emph {et~al.}(2002)\citenamefont {{Yu}},
  \citenamefont {{Kitaev}}, \citenamefont {{Shen}},\ and\ \citenamefont
  {{Vyalyi}}}]{kitaevbook}%
  \BibitemOpen
  \bibfield  {author} {\bibinfo {author} {\bibfnamefont {A.}~\bibnamefont
  {{Yu}}}, \bibinfo {author} {\bibfnamefont {A.}~\bibnamefont {{Kitaev}}},
  \bibinfo {author} {\bibfnamefont {H.}~\bibnamefont {{Shen}}}, \ and\ \bibinfo
  {author} {\bibfnamefont {M.}~\bibnamefont {{Vyalyi}}},\ }\href@noop {} {\emph
  {\bibinfo {title} {Classical and quantum computation}}},\ \bibinfo {series}
  {Graduate Studies in Mathematics}, Vol.~\bibinfo {volume} {47}\ (\bibinfo
  {publisher} {American Mathematical Society},\ \bibinfo {address} {Providence,
  RI},\ \bibinfo {year} {2002})\BibitemShut {NoStop}%
\bibitem [{\citenamefont {Bravyi}()}]{bravyi}%
  \BibitemOpen
  \bibfield  {author} {\bibinfo {author} {\bibfnamefont {S.}~\bibnamefont
  {Bravyi}},\ }\href {http://arxiv.org/abs/quant-ph/0602108v1} {\bibinfo
  {journal} {arXiv:quant-ph/0602108v1}\ }\BibitemShut {NoStop}%
\bibitem [{\citenamefont {Gosset}\ and\ \citenamefont {Nagaj}(2013)}]{gosset}%
  \BibitemOpen
\bibfield  {journal} {  }\bibfield  {author} {\bibinfo {author} {\bibfnamefont
  {D.}~\bibnamefont {Gosset}}\ and\ \bibinfo {author} {\bibfnamefont
  {D.}~\bibnamefont {Nagaj}},\ }in\ \href {\doibase 10.1109/FOCS.2013.86}
  {\emph {\bibinfo {booktitle} {Foundations of Computer Science (FOCS), 2013
  IEEE 54th Annual Symposium on}}}\ (\bibinfo {year} {2013})\ pp.\ \bibinfo
  {pages} {756--765}\BibitemShut {NoStop}%
\bibitem [{Note1()}]{Note1}%
  \BibitemOpen
  \bibinfo {note} {While both NP and QMA complete problems are believed to be
  hard for both classical and quantum computers to solve, the latter are such
  that a classical computer cannot even verify the solution
  efficiently}\BibitemShut {NoStop}%
\bibitem [{\citenamefont {{Hartmann}}\ and\ \citenamefont
  {{Weigt}}(2005)}]{weigt}%
  \BibitemOpen
  \bibfield  {author} {\bibinfo {author} {\bibfnamefont {A.}~\bibnamefont
  {{Hartmann}}}\ and\ \bibinfo {author} {\bibfnamefont {M.}~\bibnamefont
  {{Weigt}}},\ }\href@noop {} {\emph {\bibinfo {title} {Phase Transitions in
  Combinatorial Optimization Problems}}}\ (\bibinfo  {publisher} {Wiley-{VCH}
  Verlag {GmbH} and Co. {KGaA}},\ \bibinfo {address} {Weinheim},\ \bibinfo
  {year} {2005})\BibitemShut {NoStop}%
\bibitem [{\citenamefont {{Fu}}\ and\ \citenamefont
  {{Anderson}}(1989)}]{anderson}%
  \BibitemOpen
  \bibfield  {author} {\bibinfo {author} {\bibfnamefont {Y.-T.}\ \bibnamefont
  {{Fu}}}\ and\ \bibinfo {author} {\bibfnamefont {P.}~\bibnamefont
  {{Anderson}}},\ }\href@noop {} {\emph {\bibinfo {title} {Lectures in the
  Sciences of Complexity}}}\ (\bibinfo  {publisher} {Addison-Wesley},\ \bibinfo
  {address} {Reading, MA},\ \bibinfo {year} {1989})\BibitemShut {NoStop}%
\bibitem [{\citenamefont {{M\'{e}zard}}\ and\ \citenamefont
  {{Montanari}}(2009)}]{montanaribook}%
  \BibitemOpen
  \bibfield  {author} {\bibinfo {author} {\bibfnamefont {M.}~\bibnamefont
  {{M\'{e}zard}}}\ and\ \bibinfo {author} {\bibfnamefont {A.}~\bibnamefont
  {{Montanari}}},\ }\href@noop {} {\emph {\bibinfo {title} {Information,
  Physics, and Computation}}}\ (\bibinfo  {publisher} {Oxford Graduate Texts},\
  \bibinfo {address} {New York},\ \bibinfo {year} {2009})\BibitemShut {NoStop}%
\bibitem [{\citenamefont {{Kirkpatrick}}\ and\ \citenamefont
  {{Selman}}(1994)}]{kirkpatrick2}%
  \BibitemOpen
  \bibfield  {author} {\bibinfo {author} {\bibfnamefont {S.}~\bibnamefont
  {{Kirkpatrick}}}\ and\ \bibinfo {author} {\bibfnamefont {B.}~\bibnamefont
  {{Selman}}},\ }\href {\doibase 10.1126/science.264.5163.1297} {\bibfield
  {journal} {\bibinfo  {journal} {Science}\ }\textbf {\bibinfo {volume}
  {264}},\ \bibinfo {pages} {1297} (\bibinfo {year} {1994})}\BibitemShut
  {NoStop}%
\bibitem [{\citenamefont {Laumann}\ \emph {et~al.}(2009)\citenamefont
  {Laumann}, \citenamefont {Moessner}, \citenamefont {Scardicchio},\ and\
  \citenamefont {Sondhi}}]{laumann2}%
  \BibitemOpen
  \bibfield  {author} {\bibinfo {author} {\bibfnamefont {C.~R.}\ \bibnamefont
  {Laumann}}, \bibinfo {author} {\bibfnamefont {R.}~\bibnamefont {Moessner}},
  \bibinfo {author} {\bibfnamefont {A.}~\bibnamefont {Scardicchio}}, \ and\
  \bibinfo {author} {\bibfnamefont {S.~L.}\ \bibnamefont {Sondhi}},\ }\href
  {http://arxiv.org/abs/0903.1904v1} {\bibfield  {journal} {\bibinfo  {journal}
  {Quant. Inf. and Comp.}\ }\textbf {\bibinfo {volume} {10}},\ \bibinfo {pages}
  {0001} (\bibinfo {year} {2009})}\BibitemShut {NoStop}%
\bibitem [{\citenamefont {Laumann}\ \emph {et~al.}(2010)\citenamefont
  {Laumann}, \citenamefont {L\"auchli}, \citenamefont {Moessner}, \citenamefont
  {Scardicchio},\ and\ \citenamefont {Sondhi}}]{laumann3}%
  \BibitemOpen
  \bibfield  {author} {\bibinfo {author} {\bibfnamefont {C.}~\bibnamefont
  {Laumann}}, \bibinfo {author} {\bibfnamefont {A.}~\bibnamefont {L\"auchli}},
  \bibinfo {author} {\bibfnamefont {R.}~\bibnamefont {Moessner}}, \bibinfo
  {author} {\bibfnamefont {A.}~\bibnamefont {Scardicchio}}, \ and\ \bibinfo
  {author} {\bibfnamefont {S.}~\bibnamefont {Sondhi}},\ }\href {\doibase
  10.1103/PhysRevA.81.062345} {\bibfield  {journal} {\bibinfo  {journal}
  {Physical Review A}\ }\textbf {\bibinfo {volume} {81}} (\bibinfo {year}
  {2010}),\ 10.1103/PhysRevA.81.062345}\BibitemShut {NoStop}%
\bibitem [{\citenamefont {Hsu}\ \emph {et~al.}(2013)\citenamefont {Hsu},
  \citenamefont {Laumann}, \citenamefont {L\"{a}uchli}, \citenamefont
  {Moessner},\ and\ \citenamefont {Sondhi}}]{hsu}%
  \BibitemOpen
  \bibfield  {author} {\bibinfo {author} {\bibfnamefont {B.}~\bibnamefont
  {Hsu}}, \bibinfo {author} {\bibfnamefont {C.~R.}\ \bibnamefont {Laumann}},
  \bibinfo {author} {\bibfnamefont {A.~M.}\ \bibnamefont {L\"{a}uchli}},
  \bibinfo {author} {\bibfnamefont {R.}~\bibnamefont {Moessner}}, \ and\
  \bibinfo {author} {\bibfnamefont {S.~L.}\ \bibnamefont {Sondhi}},\ }\href
  {\doibase 10.1103/PhysRevA.87.062334} {\bibfield  {journal} {\bibinfo
  {journal} {Physical Review A}\ }\textbf {\bibinfo {volume} {87}} (\bibinfo
  {year} {2013}),\ 10.1103/PhysRevA.87.062334}\BibitemShut {NoStop}%
\bibitem [{\citenamefont {Arora}\ \emph {et~al.}(1998)\citenamefont {Arora},
  \citenamefont {Lund}, \citenamefont {Motwani}, \citenamefont {Sudan},\ and\
  \citenamefont {Szegedy}}]{PCPthm}%
  \BibitemOpen
  \bibfield  {author} {\bibinfo {author} {\bibfnamefont {S.}~\bibnamefont
  {Arora}}, \bibinfo {author} {\bibfnamefont {C.}~\bibnamefont {Lund}},
  \bibinfo {author} {\bibfnamefont {R.}~\bibnamefont {Motwani}}, \bibinfo
  {author} {\bibfnamefont {M.}~\bibnamefont {Sudan}}, \ and\ \bibinfo {author}
  {\bibfnamefont {M.}~\bibnamefont {Szegedy}},\ }\href {\doibase
  10.1145/278298.278306} {\bibfield  {journal} {\bibinfo  {journal} {J. ACM}\
  }\textbf {\bibinfo {volume} {45}},\ \bibinfo {pages} {501} (\bibinfo {year}
  {1998})}\BibitemShut {NoStop}%
\bibitem [{Note2()}]{Note2}%
  \BibitemOpen
  \bibinfo {note} {The ground state energy is at least a promised gap $\Delta $
  above zero if it is not precisely zero.}\BibitemShut {Stop}%
\bibitem [{\citenamefont {Schmidt-Pruzan}\ and\ \citenamefont
  {Shamir}(1985)}]{hypergraphs}%
  \BibitemOpen
  \bibfield  {author} {\bibinfo {author} {\bibfnamefont {J.}~\bibnamefont
  {Schmidt-Pruzan}}\ and\ \bibinfo {author} {\bibfnamefont {E.}~\bibnamefont
  {Shamir}},\ }\href {\doibase 10.1007/BF02579445} {\bibfield  {journal}
  {\bibinfo  {journal} {Combinatorica}\ }\textbf {\bibinfo {volume} {5}},\
  \bibinfo {pages} {81} (\bibinfo {year} {1985})}\BibitemShut {NoStop}%
\bibitem [{\citenamefont {{Erd\H{o}s}}\ and\ \citenamefont
  {{R\'{e}nyi}}(1960)}]{erdos}%
  \BibitemOpen
  \bibfield  {author} {\bibinfo {author} {\bibfnamefont {P.}~\bibnamefont
  {{Erd\H{o}s}}}\ and\ \bibinfo {author} {\bibfnamefont {A.}~\bibnamefont
  {{R\'{e}nyi}}},\ }\href {http://www.renyi.hu/~p_erdos/1960-10.pdf} {\bibfield
   {journal} {\bibinfo  {journal} {Magyar {Tud.} {Akad.} {Mat.} {Kutat{\'o}}
  {Int.} {K{\"o}zl}}\ }\textbf {\bibinfo {volume} {5}},\ \bibinfo {pages} {17}
  (\bibinfo {year} {1960})}\BibitemShut {NoStop}%
\bibitem [{\citenamefont {Bollob\'{a}s}(1998)}]{bollobas}%
  \BibitemOpen
  \bibfield  {author} {\bibinfo {author} {\bibfnamefont {B.}~\bibnamefont
  {Bollob\'{a}s}},\ }\href@noop {} {\emph {\bibinfo {title} {Modern Graph
  Theory}}}\ (\bibinfo  {publisher} {Springer},\ \bibinfo {address} {New
  York},\ \bibinfo {year} {1998})\BibitemShut {NoStop}%
\bibitem [{\citenamefont {Bravyi}\ \emph {et~al.}()\citenamefont {Bravyi},
  \citenamefont {Moore},\ and\ \citenamefont {Russell}}]{bravyimoore}%
  \BibitemOpen
  \bibfield  {author} {\bibinfo {author} {\bibfnamefont {S.}~\bibnamefont
  {Bravyi}}, \bibinfo {author} {\bibfnamefont {C.}~\bibnamefont {Moore}}, \
  and\ \bibinfo {author} {\bibfnamefont {A.}~\bibnamefont {Russell}},\ }\href
  {http://arxiv.org/abs/0907.1297} {\bibinfo  {journal} {{arXiv}:0907.1297}\
  }\BibitemShut {NoStop}%
\bibitem [{\citenamefont {de~Beaudrap}()}]{beaudrap}%
  \BibitemOpen
\bibfield  {journal} {  }\bibfield  {author} {\bibinfo {author} {\bibfnamefont
  {N.}~\bibnamefont {de~Beaudrap}},\ }\href {http://arxiv.org/abs/1403.1588}
  {\bibinfo  {journal} {arXiv:1403.1588}\ }\BibitemShut {NoStop}%
\bibitem [{\citenamefont {Ambainis}\ \emph {et~al.}(2012)\citenamefont
  {Ambainis}, \citenamefont {Kempe},\ and\ \citenamefont
  {Sattath}}]{Ambainis:2012aa}%
  \BibitemOpen
\bibfield  {journal} {  }\bibfield  {author} {\bibinfo {author} {\bibfnamefont
  {A.}~\bibnamefont {Ambainis}}, \bibinfo {author} {\bibfnamefont
  {J.}~\bibnamefont {Kempe}}, \ and\ \bibinfo {author} {\bibfnamefont
  {O.}~\bibnamefont {Sattath}},\ }\href {\doibase 10.1145/2371656.2371659}
  {\bibfield  {journal} {\bibinfo  {journal} {J. ACM}\ }\textbf {\bibinfo
  {volume} {59}},\ \bibinfo {pages} {24:1} (\bibinfo {year}
  {2012})}\BibitemShut {NoStop}%
\bibitem [{\citenamefont {{Goerdt}}(1996)}]{goerdt}%
  \BibitemOpen
  \bibfield  {author} {\bibinfo {author} {\bibfnamefont {A.}~\bibnamefont
  {{Goerdt}}},\ }\href {\doibase doi:10.1006/jcss.1996.0081} {\bibfield
  {journal} {\bibinfo  {journal} {Journal of {Computer} and {System}
  {Sciences}}\ }\textbf {\bibinfo {volume} {53}},\ \bibinfo {pages} {469}
  (\bibinfo {year} {1996})}\BibitemShut {NoStop}%
\bibitem [{\citenamefont {{Friedgut}}\ and\ \citenamefont
  {{Bourgain}}(1999)}]{friedgut}%
  \BibitemOpen
  \bibfield  {author} {\bibinfo {author} {\bibfnamefont {E.}~\bibnamefont
  {{Friedgut}}}\ and\ \bibinfo {author} {\bibfnamefont {J.}~\bibnamefont
  {{Bourgain}}},\ }\href {\doibase 10.1090/S0894-0347-99-00305-7} {\bibfield
  {journal} {\bibinfo  {journal} {Journal of the {American} {Mathematical}
  {Society}}\ }\textbf {\bibinfo {volume} {12}},\ \bibinfo {pages} {1017}
  (\bibinfo {year} {1999})}\BibitemShut {NoStop}%
\bibitem [{\citenamefont {{Bollob\'{a}s}}\ \emph {et~al.}(2001)\citenamefont
  {{Bollob\'{a}s}}, \citenamefont {{Borgs}}, \citenamefont {{Chayes}},
  \citenamefont {{Kim}},\ and\ \citenamefont {{Wilson}}}]{bollobascaling}%
  \BibitemOpen
  \bibfield  {author} {\bibinfo {author} {\bibfnamefont {B.}~\bibnamefont
  {{Bollob\'{a}s}}}, \bibinfo {author} {\bibfnamefont {C.}~\bibnamefont
  {{Borgs}}}, \bibinfo {author} {\bibfnamefont {J.}~\bibnamefont {{Chayes}}},
  \bibinfo {author} {\bibfnamefont {J.}~\bibnamefont {{Kim}}}, \ and\ \bibinfo
  {author} {\bibfnamefont {D.}~\bibnamefont {{Wilson}}},\ }\href {\doibase
  10.1002/rsa.1006} {\bibfield  {journal} {\bibinfo  {journal} {Random
  {Structures} and {Algorithms}}\ }\textbf {\bibinfo {volume} {18}},\ \bibinfo
  {pages} {201} (\bibinfo {year} {2001})}\BibitemShut {NoStop}%
\bibitem [{\citenamefont {M\'{e}zard}\ \emph {et~al.}(2003)\citenamefont
  {M\'{e}zard}, \citenamefont {Ricci-Tersenghi},\ and\ \citenamefont
  {Zecchina}}]{mezard}%
  \BibitemOpen
  \bibfield  {author} {\bibinfo {author} {\bibfnamefont {M.}~\bibnamefont
  {M\'{e}zard}}, \bibinfo {author} {\bibfnamefont {F.}~\bibnamefont
  {Ricci-Tersenghi}}, \ and\ \bibinfo {author} {\bibfnamefont {R.}~\bibnamefont
  {Zecchina}},\ }\href {\doibase 10.1023/A:1022886412117} {\bibfield  {journal}
  {\bibinfo  {journal} {Journal of Statistical Physics}\ }\textbf {\bibinfo
  {volume} {111}},\ \bibinfo {pages} {505} (\bibinfo {year}
  {2003})}\BibitemShut {NoStop}%
\bibitem [{Note3()}]{Note3}%
  \BibitemOpen
  \bibinfo {note} {We use motif to mean a fixed subgraph up to isomorphisms,
  irrespective of its frequency of occurrence. This is midway between the
  everyday meaning of the term and the more technical graph theoretic usage
  which makes reference to its frequency}\BibitemShut {NoStop}%
\bibitem [{\citenamefont {{Bollob\'{a}s}}(1984)}]{bollobasevolution}%
  \BibitemOpen
  \bibfield  {author} {\bibinfo {author} {\bibfnamefont {B.}~\bibnamefont
  {{Bollob\'{a}s}}},\ }\href {\doibase 10.1090/S0002-9947-1984-0756039-5}
  {\bibfield  {journal} {\bibinfo  {journal} {Transactions of the {American}
  {Mathematical} {Society}}\ }\textbf {\bibinfo {volume} {286}},\ \bibinfo
  {pages} {257} (\bibinfo {year} {1984})}\BibitemShut {NoStop}%
\bibitem [{\citenamefont {{{\L}uczak}}(1990)}]{luczak}%
  \BibitemOpen
  \bibfield  {author} {\bibinfo {author} {\bibfnamefont {T.}~\bibnamefont
  {{{\L}uczak}}},\ }\href {\doibase 10.1002/rsa.3240010305} {\bibfield
  {journal} {\bibinfo  {journal} {Random {Structures} and {Algorithms}}\
  }\textbf {\bibinfo {volume} {1}},\ \bibinfo {pages} {287} (\bibinfo {year}
  {1990})}\BibitemShut {NoStop}%
\end{thebibliography}%

\end{document}